%% file: conference_101719.tex
\documentclass[conference]{IEEEtran}
\IEEEoverridecommandlockouts
\usepackage{cite}
\usepackage{amsmath,amssymb,amsfonts}
\usepackage{algorithmic}
\usepackage{graphicx}
\usepackage{textcomp}
\usepackage{xcolor}
\usepackage{comment}
\usepackage{subcaption}
\usepackage{makecell}
\usepackage{multirow}
\def\BibTeX{{\rm B\kern-.05em{\sc i\kern-.025em b}\kern-.08em
    T\kern-.1667em\lower.7ex\hbox{E}\kern-.125emX}}
\begin{document}

\title{Hierarchical Quantum Control Gates for Functional MRI Understanding 
}
\newcommand{\Bac}[1]{\textcolor{cyan}{\\ \noindent Bac: #1}}
\newcommand{\ketbra}[2]{|#1\rangle\!\langle#2|}
\newcommand{\floor}[1]{\left\lfloor #1 \right\rfloor}
\newcommand{\ceil}[1]{\left\lceil #1 \right\rceil}

\author{Xuan-Bac Nguyen$^{1,*}$, Hoang-Quan Nguyen$^{1,*}$, Hugh Churchill$^{2,*}$, Samee U. Khan$^{3}$, Khoa Luu$^{1,*}$\\
    $^{1}$ CVIU Lab, University of Arkansas, AR 72703 \quad 
    $^{2}$ Dept. of Physics,  University of Arkansas, AR 72703 \\
    $^{3}$ Mississippi State University, MS 39762 \\
    $^{*}$ MonArk NSF Quantum Foundry \\
	\tt\small \{xnguyen,hn016,hchurch,khoaluu\}@uark.edu, \tt\small
        \tt\small skhan@ece.msstate.edu, \\
    
}

\maketitle

\begin{abstract}

Quantum computing has emerged as a powerful tool for solving complex problems intractable for classical computers, particularly in popular fields such as cryptography, optimization, and neurocomputing. In this paper, we present a new quantum-based approach named the Hierarchical Quantum Control Gates (HQCG) method for efficient understanding of Functional Magnetic Resonance Imaging (fMRI) data. This approach includes two novel modules: the Local Quantum Control Gate (LQCG) and the Global Quantum Control Gate (GQCG), which are designed to extract local and global features of fMRI signals, respectively. Our method operates end-to-end on a quantum machine, leveraging quantum mechanics to learn patterns within extremely high-dimensional fMRI signals, such as 30,000 samples—a challenge for classical computers. Empirical results demonstrate that our approach significantly outperforms classical methods. Additionally, we found that the proposed quantum model is more stable and less prone to overfitting than the classical methods.

\end{abstract}

\begin{IEEEkeywords}
Quantum Control Gates, Quantum Computer, fMRI, Brain-inspired Representation
\end{IEEEkeywords}

\section{Introduction}
Brain-inspired representations provide a promising pathway for enhancing the training of deep neural networks (DNN) \cite{palazzo2020decoding}. To understand these representations, it is crucial to monitor the activities of every single neuron in the brain simultaneously. Among several neurocomputing techniques such as functional Magnetic Resonance Imaging (fMRI), Magnetoencephalography (MEG), and electroencephalogram (EEG), due to the efficiency, it is worth mentioning fMRI signals that have been used widely for that purpose in most recent studies \cite{haynes2005predicting, thirion2006inverse, kamitani2005decoding, cox2003functional, haxby2001distributed, chen2023seeing, scotti2023reconstructing, takagi2023high, ozcelik2023natural, lin2022mind, ozcelik2023brain,nguyen2023algonauts,nguyen2024bractive,nguyen2023brainformer}. 
These studies explored patterns of fMRI signals in a trivial approach, i.e., using a fully connected layer to extract the features. Meanwhile, understanding the pattern of fMRI signals aims to discover the relationship between neurons that activate together. For instance, when a human perceives a visual science that contains the face of someone, the voxels within \textit{floc-face} region inside the brain will be fired or activated. For that reason, the previous approaches fall apart in exploring fMRI patterns. 

Recently, the self-attention mechanism \cite{vaswani2017attention} has been a well-known approach to Natural Language Processing (NLP) and has been widely applied across research fields such as computer vision, signal processing, etc \cite{nguyen2021clusformer,nguyen2023micron,nguyen2024insect,nguyen2020self, nguyen2022two, nguyen2019sketch, nguyen2023fairness}. 
This mechanism allows a neural network to focus on different parts of the input sequence when processing each element. This mechanism enables the network to weigh the importance of different tokens, i.e., words, subwords, etc., in the input sequence differently, depending on the context. Therefore, using the self-attention mechanism for fMRI signals is a potential approach. In particular, if we treat fMRI signals as an input sequence, the self-attention mechanism automatically measures the voxels-wide correlation inside the signals. A higher correlation means a higher chance that two or more voxels will activate together. 

The limitations when using self-attention to deal with very long sequences are high memory and complexity. Especially, the complexity and memory are quadratic and square of the sequence length. Meanwhile, fMRI signals are extremely high-dimensional signals that contain thousands of voxel activations. Thus, utilizing the self-attention mechanism for understanding fMRI signals is still a big obstacle. 

\begin{figure*}[t]
    \centering
    \includegraphics[width=0.9\linewidth]{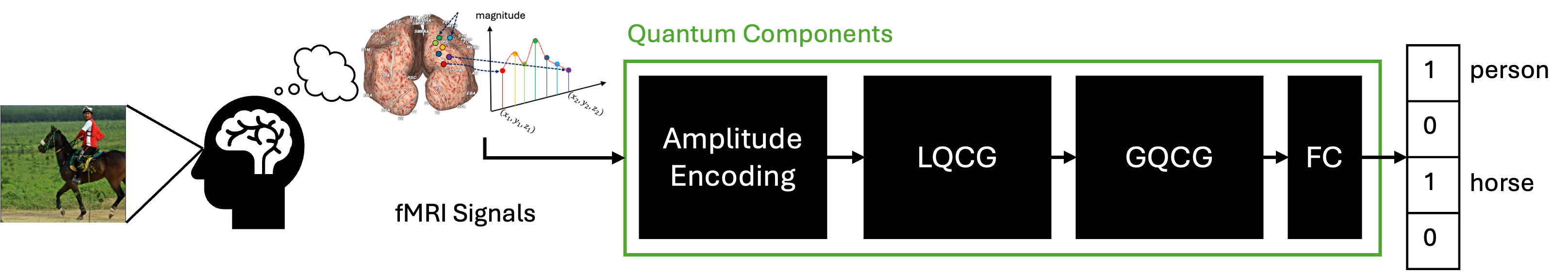}
    \caption{An overall quantum system for the Functional MRI classification. The green box includes components that run on the quantum computer: Amplitude encoding, Local Quantum Control Gate (LQCG), Global Quantum Control Gate (GQCG), and Fidelity Circuit (FC).}
    \label{fig:abstract_fig}
\vspace{-6mm}
\end{figure*}

Apart from classical computers, quantum computers can process information in parallel thanks to qubits' properties. 
Specifically, quantum bits, or qubits, can exist in a superposition of states, representing both 0 and 1 simultaneously. 
It allows a quantum computer to process a vast number of possible outcomes simultaneously, unlike classical bits, which are either 0 or 1. 
Due to superposition, a quantum computer can evaluate many possible solutions at once. 
Inspired by this property, we propose a quantum-based solution named a novel approach for fMRI understanding. 
The Fig. \ref{fig:abstract_fig} demonstrates the idea of our proposed method. The contributions in this paper can be summarized as follows:
(1) We introduce a novel quantum-based solution named Hierarchical Quantum Control Gate (HQCG) for fMRI understanding. This method can learn both local and global information of the signals in parallel. (2) Inspired by controllable gate design, we propose a trainable \textbf{local circuit} to explore the \textbf{local information} of the fMRI signal. Specifically, this module is a self-attention method that automatically learns the voxel-wise correlations. (3) To learn global information, we gather all local information extracted by \textbf{local circuits} and learn pair-wise relationships between this information by using a \textbf{global circuit}. (4) From experimental results, the proposed method performs better than a similar one running on a classical computer. Surprisingly, HQCG also helps to prevent overfitting than classical ones. 

\section{Background and Related Work}

\subsection{Brain Signal Encoding}
Decoding human brain representation has been one of the most popular research topics for a decade. In particular, cognitive neuroscience has made substantial advances in understanding neural representations originating in the primary Visual Cortex (V1) \cite{seymour2016representation}. 
Indeed, the primary visual context is in response to processing information related to oriented edges and colors. The V1 forwards the information to other neural regions, focusing more on complex shapes and features. These regions are overlapped mainly with receptive fields such as V4 \cite{peirce2015understanding}, before converging on object and category representations in the inferior temporal (IT) cortex \cite{cortex2005fast}. 
Neuroimaging techniques, including Functional Magnetic Resonance Imaging (fMRI), Magnetoencephalography (MEG), and electroencephalogram (EEG), have been crucial approaches to these studies. 
However, to replicate human-level neural representations that fully capture our visual processes, it is crucial to \textit{precisely monitor the activity of every neuron in the brain simultaneously}. 
Consequently, recent efforts in brain representation decoding have focused on exploring the correlation between neural activity data and computational models. 
In this research direction, several studies \cite{haynes2005predicting, thirion2006inverse, kamitani2005decoding, cox2003functional, haxby2001distributed} were presented to decode brain information. 
Recently, with the help of deep learning, the authors in \cite{chen2023seeing, scotti2023reconstructing, takagi2023high, ozcelik2023natural, lin2022mind, ozcelik2023brain} presented the methods to reconstruct what humans see from fMRI signal using diffusion models. 
The authors in \cite{kim2023swift, chen2023seeing} also explored patterns of fMRI signals. 
However, they often need to demonstrate or explain the nature of these patterns.
\subsection{Quantum Basics}
A quantum bit or qubit is the information carrier in the quantum computing and communication channel.
A qubit is a two-dimensional Hilbert space with two orthonormal bases $|0\rangle$ and $|1\rangle$.
These computational bases are usually represented as vectors $|0\rangle = [1, 0] ^\top$ and $|1\rangle = [0,1]^\top$.
Due to the unique qubit characteristic of superposition, the state of a qubit can be represented as the sum of two computational bases weighted by complex amplitudes as
$|\psi\rangle = \alpha |0\rangle + \beta |1\rangle$,
where $\alpha$ and $\beta \in \mathbb{C}$, and $|\alpha|^2 + |\beta|^2 = 1$. 
$|\alpha|^2$ and $|\beta|^2$ are the probability of obtaining states $|0\rangle$ and $|1\rangle$ after multiple measurements, respectively.
It gives an advantage in quantum computing over classical computing when the qubits can be entangled.
The two qubits $q_0$ and $q_1$ are entangled when they have a state that cannot be individually represented as a complex scalar times the basis vector.
A quantum state $|\psi\rangle$ can be transformed to another state $|\psi^\prime\rangle$ through a quantum circuit represented by a unitary matrix $U$.
The quantum state transformation can be mathematically formulated as $|\psi^\prime\rangle = U |\psi\rangle$.
To get classical information from a quantum state $|\psi^\prime\rangle$, quantum measurements are applied by computing the expectation value $\left<H\right> = \langle\psi^\prime| H |\psi^\prime\rangle$ of a Hermitian matrix $H$.

\subsection{Parameterized Quantum Circuit}

The parameterized quantum circuit (PQC) \cite{benedetti2019parameterized} is a unique quantum circuit with learnable parameters.
The PQC includes three modules, i.e., data encoding, parameterized layer, and quantum measurements.

Given a classical data $\mathbf{x} \in \mathbb{R}^{D}$ where $D$ is the data dimension, the data encoding circuit $U(\mathbf{x})$ is used to transform $\mathbf{x}$ into a quantum state $|\psi\rangle$.
The quantum state $|\psi\rangle$ is transformed via parameterized circuits $V(\theta)$ to a new state $|\psi\rangle$.
The parameterized circuits is a sequence of quantum circuit operators with learnable parameters denoted as:
\begin{equation}
    V(\theta) = V_L(\theta_L) V_{L-1}(\theta_{L-1}) \dots V_{1}(\theta_{1})
\end{equation}
where $L$ is the number of operators.
The quantum measurements $H$ are used to retrieve the values of the quantum state for further processing.
PQC uses a hybrid quantum-classical procedure to optimize the trainable parameters iteratively.
The popular optimization approaches include gradient descent \cite{sweke2020stochastic}, parameter-shift rule \cite{wierichs2022general,mitarai2018quantum}, and gradient-free techniques \cite{nannicini2019performance,chen2022variational}.
Despite various PQC training and inference problems \cite{mcclean2018barren,nguyen2024quantum,nguyen2024diffusion,nguyen2023quantum}, the PQC poses a potential approach for deep learning tasks in quantum settings due to its entanglement and superposition properties \cite{nguyen2024qclusformer,chen2020variational}.

\section{Proposed Approach}

\subsection{Data Encoding}
Classical computers operate with bits representing either 0 or 1. In contrast, quantum computers use quantum bits, or qubits, which can exist in superpositions of states, such as $|0\rangle$, $|1\rangle$, or any quantum superposition $\alpha |0\rangle + \beta |1\rangle$. To leverage the computational power of quantum machines, classical data must be transformed into these quantum states. Currently, there are several approaches for quantum data encoding, such as amplitude, phase, or PQC-based encoding (e.g., $U_3$). In this paper, we utilize the amplitude encoding strategy for the following reasons. First, amplitude encoding requires fewer qubits than others. Especially, given an fMRI signal length of $l$ denoted as $v = \left[v_0, \dots v_{l-1}\right] \in \mathbb{R}^{l}$, the amplitude encoding needs only $\ceil{\text{log}_2(l)}$ qubits while phase encoding requires $l$ qubits to encode whole data. Second, the amplitude encoding preserves the natural characteristic of the fMRI signals. In particular, the response of the voxel demonstrates the contribution of this voxel to the processes of the brain. For example, given $i^{th}$ and $j^{th}$ voxel, the expression of $v_i > v_j$ means the $i^{th}$ voxel is more informative than $j^{th}$ voxel. For that reason, amplitude encoding is the most suitable for encoding the fMRI signal.

\subsection{Local Quantum Control Gate}
\begin{figure}[t]
    \centering
    \includegraphics[width=0.7\linewidth]{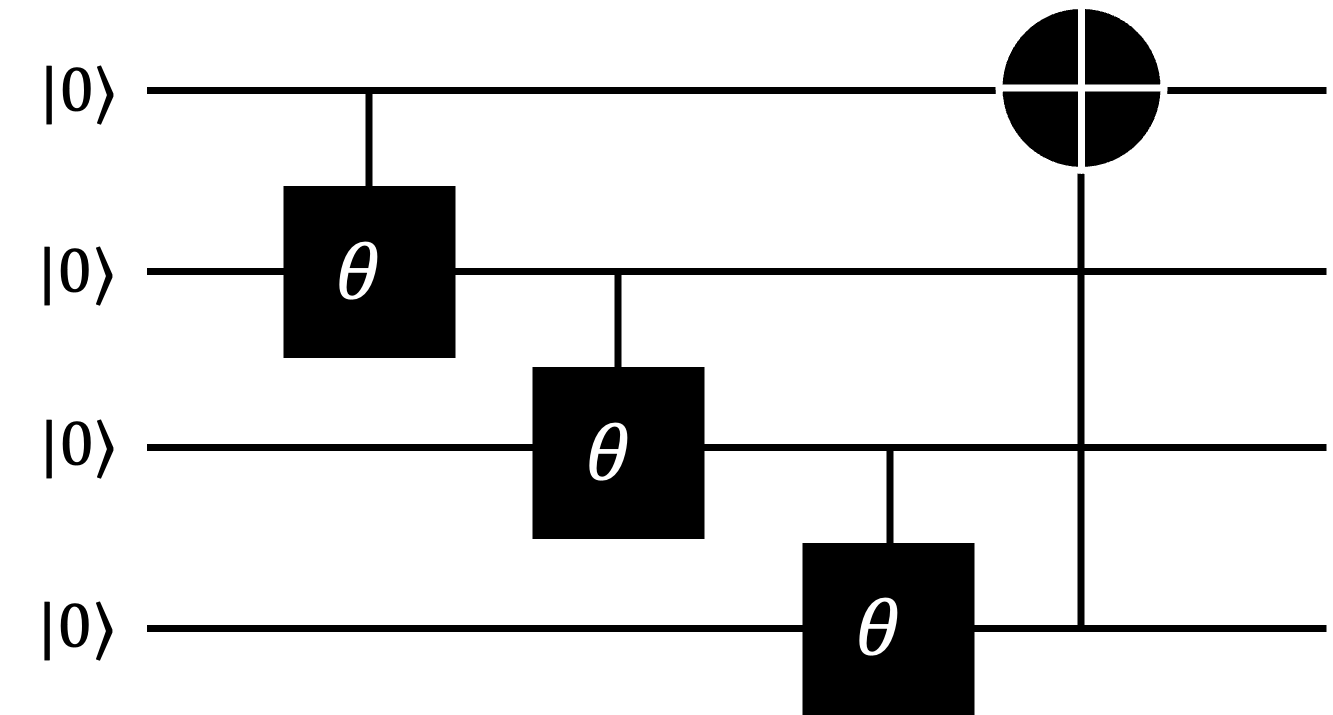}
    \caption{Local Quantum Control Gate}
    \label{fig:LQCG}
\vspace{-4mm}
\end{figure}

In fMRI signals, nearby voxels exhibit similar responses. Inspired by this observation, we propose a Local Quantum Control Gate (LQCG) to extract local features from fMRI signals. The design of LQCG is illustrated in Fig \ref{fig:LQCG}. We group continuous qubits into the LQCG, where two adjacent qubits are entangled using a trainable control unitary operation denoted as $\theta$. The output of this operation is then entangled with the next qubit. Finally, a skip connection is created by aggregating the last entangled information with the first one. The proposed design helps to combine different pieces of information, enhancing the representational power of the LQCG.
In self-attention mechanisms, each element in the sequence can attend to every other element, creating a set of attention scores that influence the representation of each element. Similarly, the entanglement in LQCG leads to non-local correlations between qubits. Therefore, the entanglement design can have a similar effect as the self-attention mechanism.

\subsection{Global Quantum Control Gate}
\begin{figure}[t]
    \centering
    \includegraphics[width=0.75\linewidth]{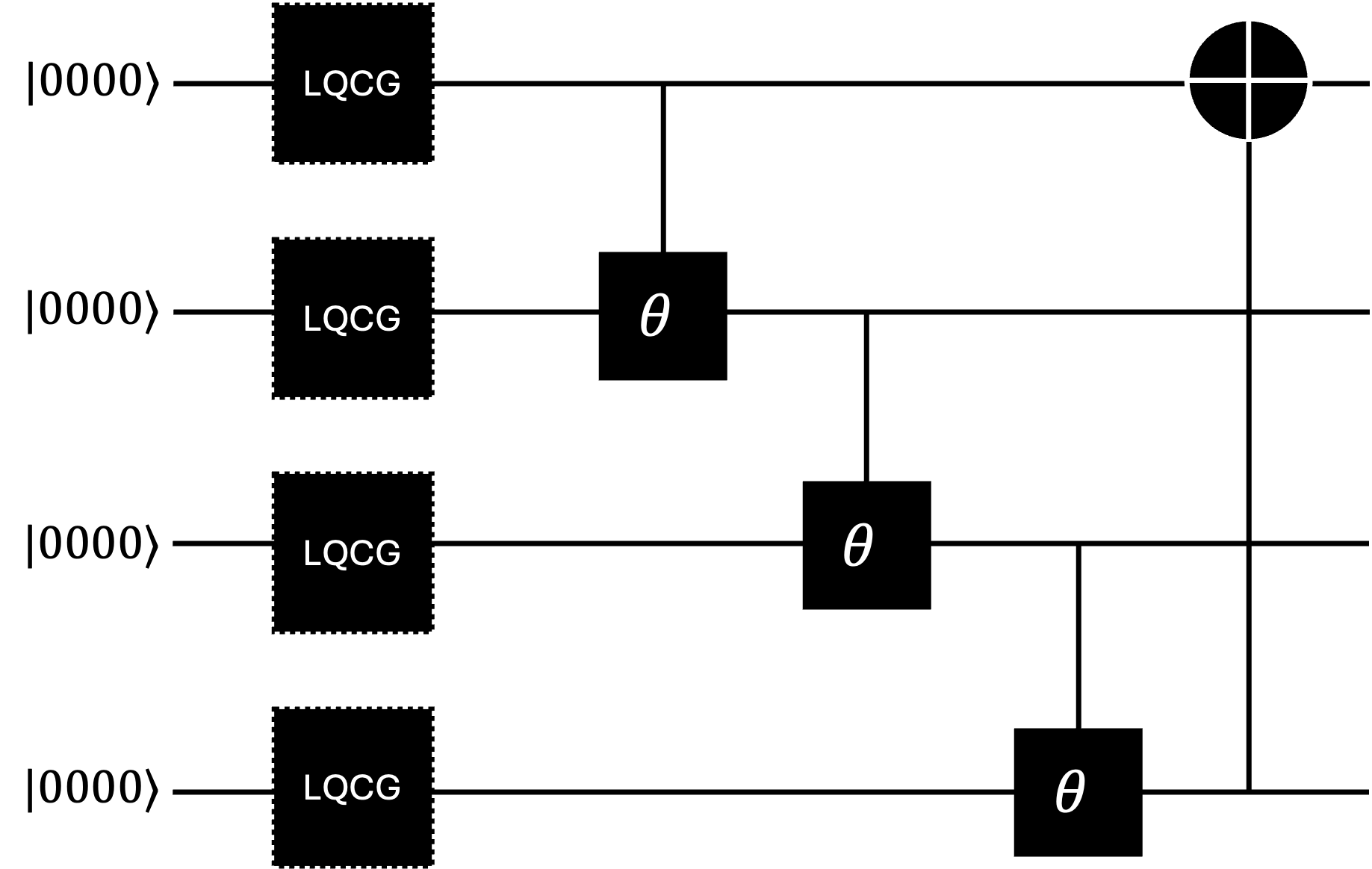}
    \caption{Global Quantum Control Gate.}
    \label{fig:GQCG}
\vspace{-4mm}
\end{figure}

As described in the previous section, we have introduced the Local Quantum Control Gate (LQCG) to extract local features from fMRI signals. To extract global features, we propose a novel approach called the Global Quantum Control Gate (GQCG). An overview of GQCG is shown in Fig \ref{fig:GQCG}. The concept of GQCG is similar to that of LQCG, incorporating hierarchical trainable control unitary operations $\theta$ and skip connections at the end. However, the critical difference is that GQCG uses the output of LQCG as its input and performs pair-wise entanglements between multiple local features. This design enables the aggregation of information from local voxels, providing a comprehensive description of the fMRI signals.

\input{tables/result}
\subsection{Multi-Classification Quantum State Fidelity Circuit}

For the classification task, we introduce a multiple quantum state fidelity circuit to classify the quantum state.
In detail, given a quantum state $|\psi\rangle$ computed from the LQCG and GQCG, we compute the fidelity of a learnable quantum state $|\phi_i\rangle$ presenting the $i$-th class.
The fidelity of the quantum states is formulated as $\langle\psi | \phi_i \rangle$.
To compute the fidelity in the quantum circuit, a swap test design is used similar to \cite{buhrman2001quantum}.

\section{Dataset and Implementation Details}
\subsection{Dataset}
We use the Natural Scenes Dataset (NSD) \cite{allen2022massive}, a comprehensive compilation of responses from eight participants obtained through high-quality 7T fMRI scans. Each subject was exposed to approximately 73,000 natural scenes, forming the basis for constructing visual brain encoding models. Since the visual stimulus in this database is a subset of COCO \cite{coco}, for each sample, visual stimulus (images) has labels of the objects inside and corresponding fMRI response, respectively. The fMRI data contains signals from both the left and right hemispheres. The length of these signals varies across subjects.

\begin{figure*}[t]
    \centering
    \includegraphics[width=0.85\linewidth]{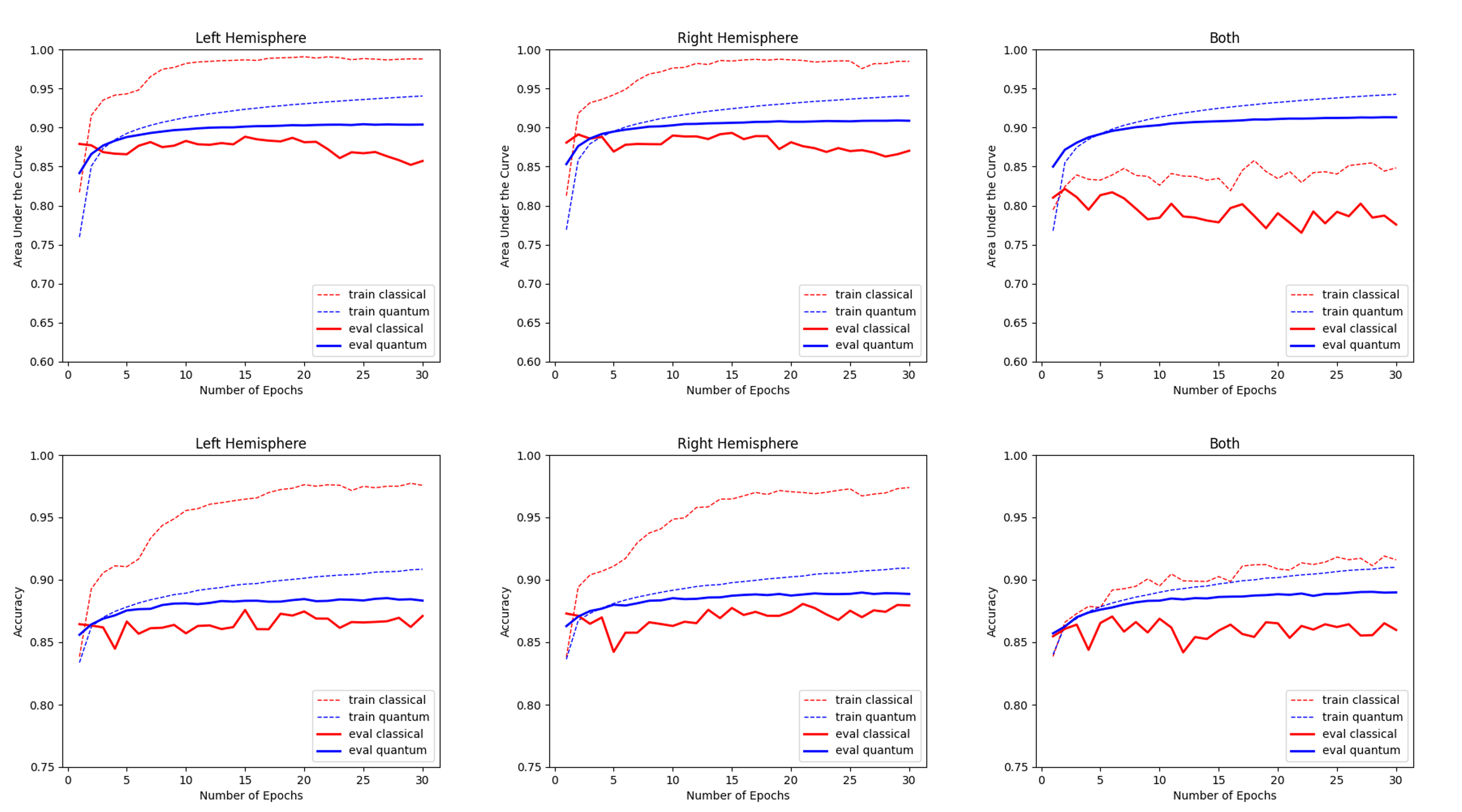}
    \caption{
    The training progresses of classical and quantum fMRI classification models using different brain hemispheres, i.e., left hemisphere, right hemisphere, and both.
    }
    \label{fig:training_plot_sides}
\vspace{-6mm}
\end{figure*}

\subsection{Implementation Details}
\noindent
\textbf{Quantum Components.}
We use the amplitude encoding method to represent the signal in the quantum system and preserve the relative magnitude of the fMRI.
The quantum system requires 16 qubits to encode and extract features from the fMRI signal.
For simplification, we use one LQCG layer and one GQCG layer.
Then, a multiple quantum state fidelity circuit is used for the multi-classification task. For a fair comparison, we employ feed-forward layers with a similar depth to the quantum components in the classical network. 

\noindent
\textbf{Objective Loss Function.} As a multi-class classification problem, we employ standard binary cross-entropy loss function to optimize the quantum networks.
\newline 
\noindent
\textbf{Training Process.} We implement the network using the TorchQuantum library \cite{hanruiwang2022quantumnas} to simulate the quantum machine. 
Since this library uses PyTorch as the backend, we can leverage GPUs to speed up the training process. 
The models are trained on an A100 GPU with 40GB of memory. The learning rate starts at $0.01$ and progressively decreases to zero following the CosineAnnealing policy \cite{loshchilov2016sgdr}. The model is trained with a batch size of $64$, AdamW \cite{loshchilov2017decoupled} optimizer for $30$ epochs, with a training time of approximately 5 minutes.
\newline
\noindent
\textbf{Evaluation Metrics.} We use accuracy and Area Under Curve (AUC) as the metrics for the comparison.

\begin{figure*}[t]
    \centering
    \includegraphics[width=0.95\linewidth]{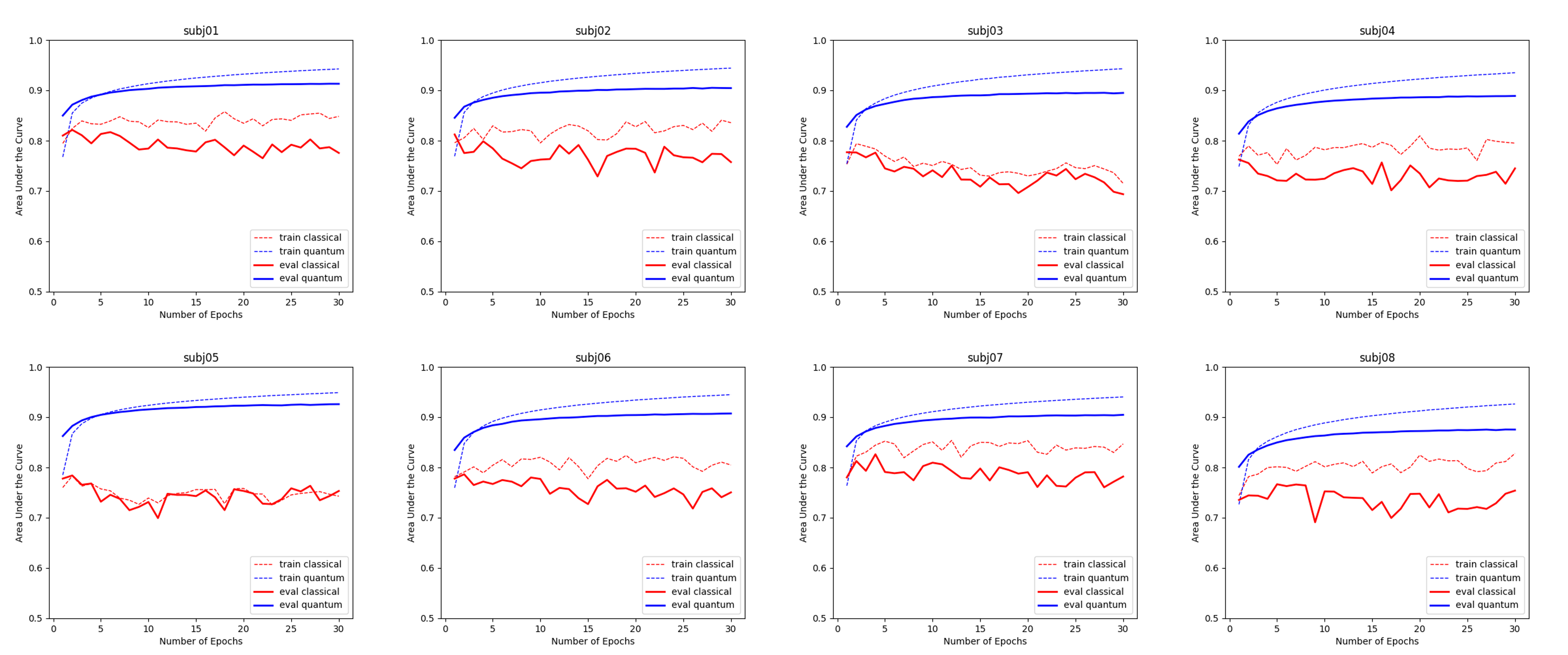}
    \caption{
    The training progresses of classical and quantum fMRI classification models on different subjects.
    }
    \label{fig:training_plot_all_subjects}
\vspace{-4mm}
\end{figure*}

\begin{figure*}[t]
    \centering
    \includegraphics[width=0.85\linewidth]{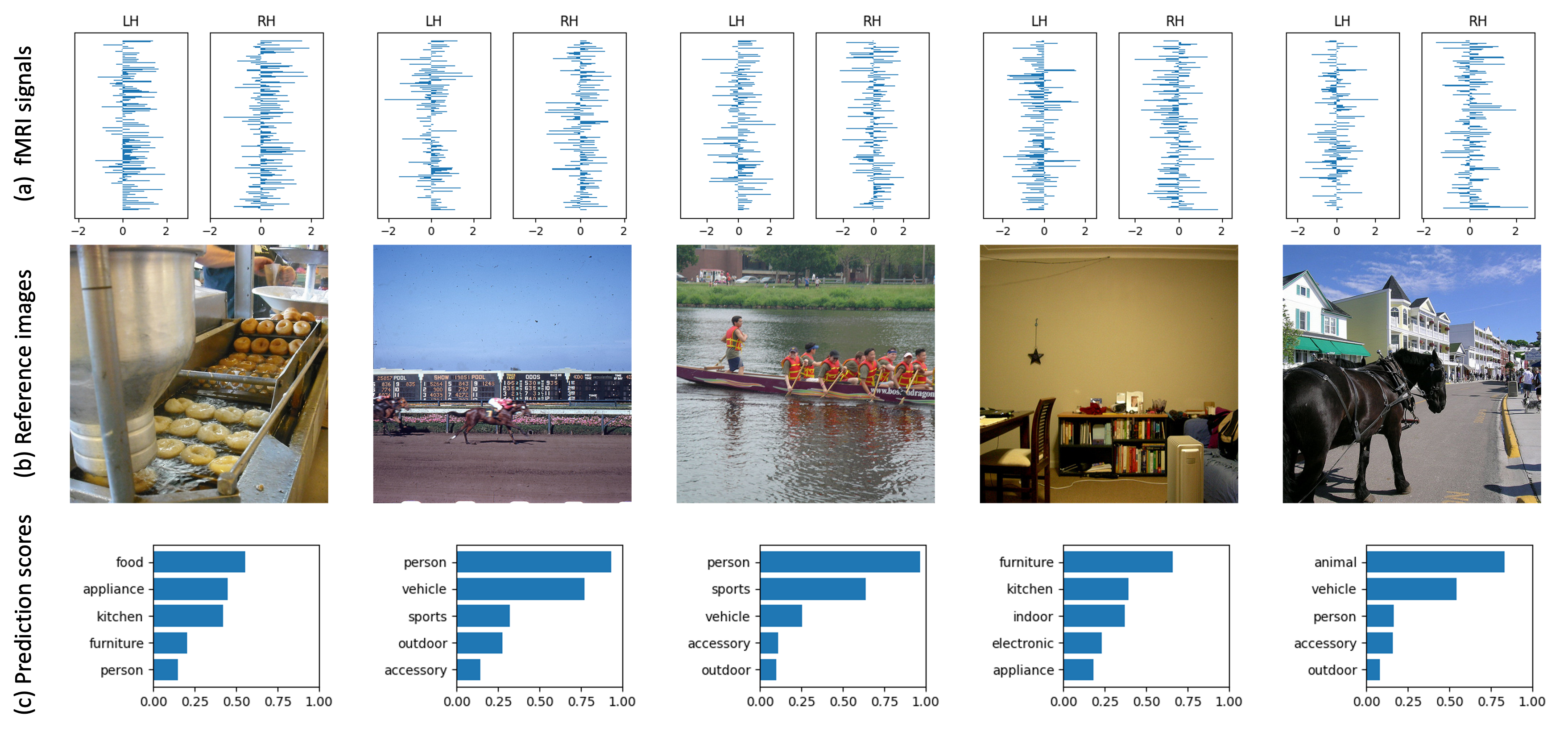}
    \caption{The prediction examples of the proposed HQCG model on the NSD dataset.}
    \label{fig:prediction_example}
\vspace{-6mm}
\end{figure*}

\section{Experiments and Results}

\subsection{Multi-Objects Predictions From fMRI}
This task aims to predict which objects a participant perceives based on recorded fMRI signals. 
The dataset includes fMRI signals from both the left and right hemispheres.
Table \ref{tab:evaluation} illustrates the performance of these individual hemispheres and their combination.

Compared to the classical approach, our proposed method achieves approximately 1\%-3\% higher accuracy for signals from both the left and right hemispheres. Notably, the quantum approach significantly outperforms the classical one when dealing with a combination of signals from both sides of the brain, with an improvement of approximately 8\%-14\%.

Interestingly, it is clear that for all subjects (from Subj01 to Subj08), the performance of the model running on a classical computer is lower when combining both hemispheres than using either the left or right hemisphere alone. Meanwhile, we observe that the quantum model maintains better performance than when using signals from just the left or right hemispheres. It can be explained by combining signals from both hemispheres, resulting in a more extended sequence, which challenges the classical approach. In contrast, the quantum approach effectively handles this more extended sequence. This result demonstrates the efficiency of our proposed method and highlights the potential of quantum computing. 

\subsection{Training Stability}

In this section, we analyze the training stability of the models on classical and quantum computing for fMRI classification problems. We report the accuracy and AUC for both training and validation during the training process, as shown in Fig \ref{fig:training_plot_sides}. Initially, when training with the left or right hemispheres, we observe that the training curves of the classical approach are significantly higher than those of the quantum approach; however, the validation curves for the classical approach are consistently lower than those of the quantum approach. It indicates that the classical approach is more prone to overfitting. Additionally, the classical method's curves exhibit more fluctuations, while the quantum method's curves are stable and smooth, demonstrating greater training stability. In conclusion, the quantum approach can prevent overfitting and stabilize the training process, leading to better results than the classical approach. Fig \ref{fig:training_plot_all_subjects} shows the training progress of classical and quantum models on different subjects. Clearly, our approach's performances and stability consistently perform well on various subjects.

\subsection{Prediction Demonstrations}
Fig \ref{fig:prediction_example} resents example outputs of our proposed method. Part (a) displays the signals from the left and right hemispheres, part (b) shows the visual stimuli that the subject is viewing, and part (c) illustrates the prediction scores. The prediction bars indicate that our method achieves significantly high confidence scores.

\section{Conclusion and Discussion}

In this paper, we present a novel quantum-based method for understanding fMRI data. This method comprises two main components, the Local Quantum Control Gate (LQCG) and the Global Quantum Control Gate (GQCG), designed to learn and extract local and global features from extremely long fMRI signals, such as 30,000 samples. Empirical experiments demonstrate the superior efficiency and stability of our approach on a quantum computer compared to its performance on a classical computer. Implemented to run end-to-end on a quantum machine, our approach leverages quantum mechanics to advance neuroscience and could inspire applications in other fields in the future.

\noindent
\textbf{Acknowledgment.} 
This work is partly supported by MonArk NSF Quantum Foundry, the National Science Foundation Q-AMASE-i program under NSF award No. DMR-1906383, the University of Arkansas Travel Grant, and JBHunt Company.

{
    \small
    \bibliographystyle{IEEEtran}
    \bibliography{main}
}

\end{document}

%% file: tables/result.tex
\begin{table*}[t]
\centering
\caption{
    Evaluation results on the NSD dataset.
    We compute the accuracy and area under the ROC curve (AUC) of the predictions on different subjects and hemispheres.
}
\resizebox{0.9\linewidth}{!}{
\begin{tabular}{cl|ccc|ccc|ccc|ccc}
\Xhline{2\arrayrulewidth}
 &
   &
  \multicolumn{3}{c}{Subj01} &
  \multicolumn{3}{c}{Subj02} &
  \multicolumn{3}{c}{Subj03} &
  \multicolumn{3}{c}{Subj04} \\
 &
   &
  LH &
  RH &
  Both &
  LH &
  RH &
  Both &
  LH &
  RH &
  Both &
  LH &
  RH &
  Both \\
\hline
\multicolumn{1}{c}{\multirow{2}{*}{Accuracy}} &
  Classical &
  87.59\% &
  88.07\% &
  87.06\% &
  87.67\% &
  86.96\% &
  86.14\% &
  87.48\% &
  87.68\% &
  87.29\% &
  86.16\% &
  86.36\% &
  85.43\% \\
\multicolumn{1}{c}{} &
  Quantum &
  \textbf{88.53\%} &
  \textbf{88.98\%} &
  \textbf{89.04\%} &
  \textbf{88.89\%} &
  \textbf{88.62\%} &
  \textbf{88.89\%} &
  \textbf{88.59\%} &
  \textbf{88.58\%} &
  \textbf{88.88\%} &
  \textbf{87.38\%} &
  \textbf{88.17\%} &
  \textbf{87.97\%} \\
\hline
\hline
 &
   &
  \multicolumn{3}{c}{Subj05} &
  \multicolumn{3}{c}{Subj06} &
  \multicolumn{3}{c}{Subj07} &
  \multicolumn{3}{c}{Subj08} \\
 &
   &
  LH &
  RH &
  Both &
  LH &
  RH &
  Both &
  LH &
  RH &
  Both &
  LH &
  RH &
  Both \\
\hline
\multicolumn{1}{c}{\multirow{2}{*}{Accuracy}} &
  Classical &
  88.43\% &
  88.66\% &
  87.42\% &
  87.75\% &
  87.71\% &
  86.70\% &
  86.71\% &
  87.28\% &
  86.86\% &
  85.44\% &
  86.11\% &
  84.94\% \\
\multicolumn{1}{c}{} &
  Quantum &
  \textbf{89.91\%} &
  \textbf{89.85\%} &
  \textbf{90.03\%} &
  \textbf{88.93\%} &
  \textbf{89.32\%} &
  \textbf{89.27\%} &
  \textbf{88.02\%} &
  \textbf{88.68\%} &
  \textbf{88.87\%} &
  \textbf{87.15\%} &
  \textbf{87.58\%} &
  \textbf{87.88\%} \\
\hline
\hline
 &
   &
  \multicolumn{3}{c}{Subj01} &
  \multicolumn{3}{c}{Subj02} &
  \multicolumn{3}{c}{Subj03} &
  \multicolumn{3}{c}{Subj04} \\
 &
   &
  LH &
  RH &
  Both &
  LH &
  RH &
  Both &
  LH &
  RH &
  Both &
  LH &
  RH &
  Both \\
\hline
\multicolumn{1}{c}{\multirow{2}{*}{AUC}} &
  Classical &
  88.82\% &
  89.30\% &
  82.14\% &
  88.76\% &
  87.40\% &
  81.23\% &
  86.93\% &
  86.85\% &
  77.70\% &
  85.27\% &
  86.11\% &
  76.25\% \\
\multicolumn{1}{c}{} &
  Quantum &
  \textbf{90.42\%} &
  \textbf{90.89\%} &
  \textbf{91.31\%} &
  \textbf{90.34\%} &
  \textbf{89.52\%} &
  \textbf{90.50\%} &
  \textbf{88.86\%} &
  \textbf{88.96\%} &
  \textbf{89.51\%} &
  \textbf{87.72\%} &
  \textbf{89.00\%} &
  \textbf{88.89\%} \\
\hline
\hline
 &
   &
  \multicolumn{3}{c}{Subj05} &
  \multicolumn{3}{c}{Subj06} &
  \multicolumn{3}{c}{Subj07} &
  \multicolumn{3}{c}{Subj08} \\
 &
   &
  LH &
  RH &
  Both &
  LH &
  RH &
  Both &
  LH &
  RH &
  Both &
  LH &
  RH &
  Both \\
\hline
\multicolumn{1}{c}{\multirow{2}{*}{AUC}} &
  Classical &
  90.20\% &
  90.52\% &
  78.40\% &
  87.50\% &
  88.22\% &
  78.64\% &
  87.31\% &
  88.28\% &
  82.62\% &
  83.84\% &
  84.32\% &
  76.66\% \\
\multicolumn{1}{c}{} &
  Quantum &
  \textbf{92.16\%} &
  \textbf{92.15\%} &
  \textbf{92.58\%} &
  \textbf{89.81\%} &
  \textbf{90.40\%} &
  \textbf{90.73\%} &
  \textbf{89.24\%} &
  \textbf{90.14\%} &
  \textbf{90.47\%} &
  \textbf{86.43\%} &
  \textbf{87.17\%} &
  \textbf{87.55\%} \\
\Xhline{2\arrayrulewidth}
\end{tabular}
}
\label{tab:evaluation}
\vspace{-4mm}
\end{table*}